\documentclass[a4paper,fleqn,usenatbib]{mnras}

\usepackage[T1]{fontenc}
\usepackage{ae,aecompl}

\usepackage{graphicx}	
\usepackage{amsmath}	
\usepackage{amssymb}	
\usepackage{gensymb}

\defcitealias{santos-lima_etal_2014}{SL+14}
\defcitealias{nakwacki_etal_2016}{Paper I}

\title[Statistics from simulated Faraday rotation maps of the intracluster medium]{Features of collisionless turbulence in the intracluster \\
medium from simulated Faraday rotation maps II:  \\ 
the effects of instabilities feedback}

\author[R. Santos-Lima al.]{R. Santos-Lima,$^{1}$\thanks{E-mail: rlima@astro.iag.usp.br}
E.~M. de Gouveia Dal Pino,$^{1}$
D.~A. Falceta-Gon\c calves,$^{2}$
\newauthor 
M.~S. Nakwacki,$^{3,4}$
and G. Kowal$^{5, 2}$
\\
$^{1}$Instituto de Astronomia, Geof\'isica e Ci\^encias Atmosf\'ericas, 
Universidade de S\~ao Paulo, R. do Mat\~ao, 1226, S\~ao Paulo, SP 05508-090, Brazil\\
$^{2}$Escola de Artes, Ci\^encias e Humanidades, Universidade de S\~ao Paulo, S\~ao Paulo, SP, Brazil\\
$^{3}$Instituto de Astronom\'ia y F\'isica del Espacio, UBA-CONICET, Argentina\\
$^{4}$Facultad de Ciencias Exactas y Naturales, Universidad de Buenos Aires, Argentina\\
$^{5}$N\'ucleo de Astrof\'isica Te\'orica, Universidade Cruzeiro do Sul, S\~ao Paulo, SP, Brazil
}

\date{Accepted XXX. Received YYY; in original form ZZZ}

\pubyear{2016}

\begin{document}
\label{firstpage}
\pagerange{\pageref{firstpage}--\pageref{lastpage}}
\maketitle

\begin{abstract}

Statistical analysis of Faraday Rotation Measure (RM) maps of the intracluster medium (ICM) of galaxy clusters 
provides a unique tool to evaluate some spatial features of the magnetic fields there.
Its combination with numerical simulations of magnetohydrodynamic (MHD)
turbulence allows the diagnosis of the ICM turbulence.
Being the ICM plasma weakly collisional, the thermal velocity distribution of the particles 
naturally develops anisotropies as a consequence 
of the large scale motions and the conservation of the magnetic moment of the charged particles.
A previous study (Paper I) analyzed the impact of large scale thermal anisotropy on the statistics of RM maps synthesized from 
simulations of turbulence; these simulations employed a collisionless MHD model which considered a tensor pressure with
uniform anisotropy.
In the present work, we extend that analysis 
to a collisionless MHD model in which the thermal anisotropy develops according 
to the conservation of the magnetic moment of the thermal particles. 
We also consider the effect 
of anisotropy relaxation caused by the micro-scale mirror and firehose instabilities. 
We show that if the relaxation rate is fast enough to keep the anisotropy limited by 
the threshold values of the instabilities, the dispersion and power spectrum 
of the RM maps are indistinguishable from those obtained from collisional MHD.
Otherwise, there is a reduction in the dispersion and steepening of the power spectrum of the RM maps 
(compared to the collisional case).
Considering the first scenario, 
the use of collisional MHD simulations for modeling the RM statistics in the ICM becomes better justified.

\end{abstract}




\graphicspath{{./figs/}}

\section{Introduction}

Cosmological mergers of galaxy clusters, AGN jets, galactic winds, and galaxy interactions drive 
turbulence in the plasma filling the 
intracluster medium (ICM), and this turbulence would be able to amplify weak seeds of
magnetic fields up to intensities of $\sim \mu G$, according to cosmological magneto-hydrodynamical
(MHD) simulations (e.g. \citealt{kotarba_etal_2011, beresnyak_miniati_2016, egan_etal_2016}). 
This amplification mechanism could explain 
the magnetic fields detected in the diffuse ICM through synchroton emission of relativistic 
electrons in radio halos, and also through the Faraday rotation of polarized emission from 
radio sources embedded or behind the galaxy clusters (see \citealt{brunetti_jones_2014} and references 
therein). In fact, two-point statistics of the Faraday rotation maps of the ICM 
reveal a magnetic field power spectrum consistent with a Kolmogorov-like power law $\propto k^{-5/3}$
\citep{ensslin_etal_2005}.

However, the use of the
standard MHD approximation to describe the dynamics of the ICM 
and the development of the small-scale turbulent dynamo 
is, in principle, 
not well justified
since it implies a high collisionality of the plasma particles to ensure the local thermodynamical equilibrium. 
Considering that 
the mean-free-path for the ion-ion Coulomb collisions is typically $\lambda_{ii} \sim 30$ kpc in the ICM
(considering a density $n = 10^{-3}$~cm$^{-3}$ and temperature $T=10^8$~K; see also the ion mean-free-path distribution 
inferred from cosmological simulations in \citealt{egan_etal_2016}), 
collisionless 
effects should be taken into account at least for scales $\lesssim \lambda_{ii}$ (see \citealt{schekochihin_cowley_2006}).
The most obvious
effect is the natural development of pressure (or temperature) anisotropy with respect 
to the local magnetic field. 
As a consequence there is the triggering of
electromagnetic plasma instabilities 
(such as the firehose, the ion-cyclotron and the mirror instabilities; see, e.g., \citealt{gary_1993}).
These instabilities are known to constrain the anisotropy itself 
(see \citealt{santos-lima_etal_2014} and references therein; \citealt{kunz_etal_2014, riquelme_etal_2015, 
sironi_narayan_2015, sironi_2015, rincon_etal_2015, melville_etal_2016, santos-lima_etal_2016}). 

\citet[\citetalias{santos-lima_etal_2014} hereafter]{santos-lima_etal_2014} took into account some collisionless effects in numerical simulations of turbulence
and small-scale dynamo in MHD numerical simulations considering the conditions typical of the ICM. 
They found that under forced turbulence, the perpendicular temperature to the local magnetic field dominates the parallel 
temperature in most of the system (the parallel temperature dominates only in narrow regions 
of high compression or magnetic field reversals), leading to strong modifications in the turbulence statistics
(see also \citealt{kowal_etal_2011, falceta15} for studies on collisionless turbulence with constant pressure anisotropy)
and the complete failure of the dynamo.
On the other hand, including the relaxation of the temperature anisotropy resulting from 
the microscale (scales below those resolved in the simulation, down to the ions kinetic scales) 
plasma instabilities, the system 
gradually converges to a similar behaviour to that 
obtained by collisional MHD, depending on the anisotropy relaxing rate~\footnote{It should be 
made clear that the anisotropy relaxation employed in \citetalias{santos-lima_etal_2014} 
does not drive the pressure components to the isotropic state, but to the instabilities thresholds. 
Therefore, the similarity to the collisional MHD results is not trivial.}. 
In \citetalias{santos-lima_etal_2014} it is argued that this relaxing rate is much 
faster than the MHD time-scales, and the model 
that better represents the ICM constrains the maximum anisotropy levels to values very 
close to the plasma stable regime.

The imprints of the large scale (of the order of turbulence injection scale) 
temperature anisotropy on the Faraday rotation 
maps were first studied in \citet[\citetalias{nakwacki_etal_2016} hereafter]{nakwacki_etal_2016}. 
In that work, we employed  a collisionless MHD formalism
with a double-isothermal closure (as implemented in \citealt{kowal_etal_2011}) to analyse the statistical 
properties  of the RM maps for several models of turbulence considering different values 
 of the fixed temperature anisotropy and different regimes of sub/super-Alfv\'enic and
trans/supersonic turbulence. The effects of the temperature anisotropy on the magnetic field structure and the RM maps were found to be significant evidencing  smaller
correlation lengths when compared to collisional MHD models.
In  that study it was neglected the feedback of the microscale instabilities on the plasma 
which may cause the reduction of the thermal anisotropy as described in 
\citetalias{santos-lima_etal_2014} (see also \citealt{schekochihin_cowley_2006}). 

In this work, we will extend the analysis of \citetalias{nakwacki_etal_2016} by including this effect.
We  will explore the collisionless effects on the Faraday rotation maps focusing on the turbulence 
models of the intracluster medium presented in 
\citetalias{santos-lima_etal_2014}, in which the anisotropy in temperature evolves according to the CGL closure 
\citep{chew_etal_1956} modified to include an anisotropy relaxing term. 
The important advantage of this new approach is not to use the double-isothermal closure, in which the temperature anisotropy is a fixed constant.
We will compare the RM maps and related statistical properties of two collisionless MHD models, 
one similar to the models of \citetalias{nakwacki_etal_2016} (i.e. without any anisotropy relaxation), 
and another including bounds in the anisotropy.
We will also  compare these  with the Faraday rotation maps obtained  from a standard collisional MHD model.

In Section 2 we describe the numerical simulations of the collisionless MHD models used for building the synthetic 
Faraday rotation maps, which are analysed in Section 3. In Section 4 we summarize our results and draw our conclusions.

\section{Numerical simulations}

\begin{table*}
	\centering
	\caption{Parameters and statistics of the simulations used to build the synthetic RM maps.}
	\label{tab:models}
	\begin{tabular}{lcccccccc}
		\hline
		run & 
                $B_0^2$ & 
                $c_{S0}^2$ & 
                $\langle u^2 \rangle$ & 
                $\langle B^2 \rangle$ & 
                $\langle \beta \rangle$ & 
                $\langle M_A \rangle$ & 
                res. &
                snapshots \\
		\hline
		CGL1 & 
                $0.09$ & 
                $1$ & 
                $0.59 (0.54)$ &
                $0.25 (0.35)$ &
                $17 (4.8 \times 10^2)$ &
                $1.8 (1.3)$ &
                $512^3$ &
		$4$ \\
		BA1 & 
                $0.09$ & 
                $1$ & 
                $0.48 (0.40)$ &
                $0.51 (0.33)$ &
                $16 (4.8 \times 10^2)$ &
                $1.2 (1.5)$ &
                $512^3$ &
		$4$ \\
		MHD1 & 
                $0.09$ & 
                $1$ & 
                $0.55 (0.48)$ &
                $0.58 (0.47)$ &
                $17 (9.4 \times 10^2)$ &
                $1.3 (1.5)$ &
                $512^3$ &
		$4$ \\
		\hline
		CGL2 & 
                $10^{-6}$ & 
                $1$ & 
                $0.70 (0.73)$ &
                $1.2 \times 10^{-5} (7.1 \times 10^{-5})$ &
                $2.0 \times 10^6 (4.7 \times 10^7)$ &
                $5.9 \times 10^2 (5.7 \times 10^2)$ &
                $256^3$ &
		$11$ \\
		CGL3 & 
                $10^{-6}$ & 
                $0.09$ & 
                $0.79 (0.64)$ &
                $2.0 \times 10^{-4} (7.5 \times 10^{-4})$ &
                $1.2 \times 10^5 (1.1 \times 10^7)$ &
                $2.8 \times 10^2 (4.2 \times 10^2)$ &
                $256^3$ &
		$11$ \\
		BA2 & 
                $10^{-6}$ & 
                $1$ & 
                $0.78 (0.63)$ &
                $0.12 (0.15)$ &
                $1.7 \times 10^2 (6.2 \times 10^3)$ &
                $4.6 (6.7)$ &
                $256^3$ &
		$11$ \\
		MHD2 & 
                $10^{-6}$ & 
                $1$ & 
                $0.79 (0.63)$ &
                $0.18 (0.22)$ &
                $1.1 \times 10^2 (2.5 \times 10^3)$ &
                $3.8 (5.3)$ &
                $256^3$ &
		$11$ \\
		\hline
	\end{tabular}
\end{table*}

Table~\ref{tab:models} shows the most relevant parameters of the simulated models used to 
build the synthetic RM maps. 
The brackets $\langle \cdot \rangle$ denote an average over the domain and time (using the 
available snapshots of the simulations, considering time intervals larger than 
$\tau_{turb}$, where $\tau_{turb} = L_{turb} / U_{turb}$ is the turbulence 
turn-over time, with $L_{turb}$ and $U_{turb}$ the scale and velocity of injection, respectively),  
when the turbulence has reached a statistically 
stationary state. The values listed in parenthesis are the statistical standard deviations which give an approximate idea about the spatial and temporal fluctuations of these 
quantities.~\footnote{The 
spatial/temporal statistical distributions of the fields listed in Table~\ref{tab:models} are 
not Gaussian around the mean values. This becomes obvious 
from the fact that all the quantities are positive and
some of the standard deviation values are larger
than the mean values.}
The first three models ({\it CGL1, BA1,} and {\it MHD1}) have the same initial 
uniform magnetic field with intensity $B_0$ and  thermal speed $c_{s0}$ 
(both shown in dimensionless code units; see below). The  
rms turbulent velocity is also similar in these models (column $\langle u^2 \rangle$).
The initial thermal speed is kept approximately constant 
by the use of a fast thermal relaxation (which represents the action of both radiative cooling and heat conduction; 
see more details below).
The regimes of turbulence achieved for these three models are similar 
being  slightly subsonic ($u_{rms} \lesssim c_{S0}$) and mildly super-Alfvenic ($\langle M_A \rangle \gtrsim 1$).
From all the models studied in \citetalias{santos-lima_etal_2014}, only these three ones  have simulation parameters which are similar 
to those of the super-Alfvenic simulations analysed in \citetalias{nakwacki_etal_2016} 
and thus can be more easily  compared with this previous work.
However, the ICM is  observed to have very tangled magnetic fields (e.g. \citealt{ferreti_etal_1995}), 
which is indicative of a turbulence regime strongly super-Alfvenic. 
The remaining models presented in Table~\ref{tab:models} ({\it CGL2, CGL3, BA2}, and {\it MHD2}) are 
simulations in which the 
initial uniform magnetic field is very weak, 
making the intensity of the ordered component of the magnetic field to be relatively small
after the amplification of the tangled component via  small-scale turbulent dynamo 
(see $B_0^2$ and $\langle B^2 \rangle$ for these models in Table~\ref{tab:models}).
These four models have the same initial seed magnetic field and thermal speed, except for 
model {\it CGL3}, where a smaller thermal speed is used in order to test the dependence of the results with 
the plasma $\beta = p_{th} / p_{mag}$  parameter
(where $p_{th} = (2p_{\perp} + p_{\parallel})/3$ is the total thermal pressure and $p_{mag} = B^2/8\pi$ 
is the magnetic pressure).

The models MHD ({\it MHD1} and {\it MHD2}) have 
a single scalar thermal pressure and correspond to the standard collisional MHD model where the 
distribution of the thermal velocities is assumed to be isotropic. The models named 
CGL ({\it CGL1, CGL2}, and {\it CGL3}) have
a thermal pressure tensor with two independent components related to two temperatures:
one associated to the thermal velocity component parallel to the local magnetic field lines $T_{\parallel}$ 
and another to the thermal velocity component related to the gyromotions of the particles around 
the field $T_{\perp}$; these two temperatures evolve according to the CGL closure 
\citep{chew_etal_1956}
which is based on the conservation of the magnetic moment of the charged particles 
$d \left( T_{\perp} / B \right) / dt = 0$
and the assumption of conservation of the
entropy (no heat exchange between the fluid elements)
$d \left( T_{\perp}^{2} T_{\parallel} / n^{2} \right) / dt = 0$ (where $n$ is the density of particles)~\footnote{In fact 
the CGL closure was not rigorously adopted in \citetalias{santos-lima_etal_2014}. 
Instead, a conservative scheme for evolving the internal energy was used, while the 
evolution of the temperatures ratio followed the CGL prescription (see Eq.~\ref{eqn:collisionless_mhd}). 
This approach gives results nearly 
identical to those obtained using the CGL equations of state, but is numerically more robust. Besides, 
it allows the straight inclusion of the anisotropy relaxation term.}.
Finally, the models named BA 
(Bounded Anisotropy: {\it BA1} and {\it BA2}) 
differ from  models CGL by the addition of a boundary in the temperature anisotropy. 
This boundary limits the temperature anisotropy by the threshold values of the 
firehose (for $A<1$) and mirror (for $A>1$) instabilities, where $A$ is the temperatures 
ratio $A = T_{\perp}/T_{\parallel}$, mimicking the effect of an ``instantaneous'' relaxing 
of the anisotropy to the marginal values by the action of the microscale instabilities
(see \citealt{sharma_etal_2006}; \citetalias{santos-lima_etal_2014} and references therein).
An extended discussion on the applicability and limitations of this model to represent the 
ICM turbulence is presented in \citet{santos-lima_etal_2016}.

The equations describing the evolution of the models presented in Table~\ref{tab:models} are 
(see also \citetalias{santos-lima_etal_2014}):
{\small
\begin{equation}
  \frac{\partial }{\partial t}
  \begin{bmatrix}
    \rho \\[6pt]
    \rho \mathbf{u} \\[6pt]
    \mathbf{B} \\[6pt]
    e \\[6pt]
    A (\rho^{3}/B^{3})
  \end{bmatrix}
  + \nabla \cdot
  \begin{bmatrix}
    \rho \mathbf{u} \\[6pt]
    \rho \mathbf{uu} + \Pi_{P} + \Pi_{B} \\[6pt]
    \mathbf{uB - Bu} \\[6pt]
    e \mathbf{u} + \mathbf{u} \cdot \left( \Pi_{P} + \Pi_{B} \right) \\[6pt]
    A (\rho^{3}/B^{3}) \mathbf{u}
  \end{bmatrix}
  =
  \begin{bmatrix}
    0 \\[6pt]
    \mathbf{f} \\[6pt]
    0 \\[6pt]
    \mathbf{f \cdot v}  + \dot{w} \\[6pt]
    \dot{A}_{S} (\rho^{3}/B^{3})
  \end{bmatrix}
  \rm{,}
\label{eqn:collisionless_mhd}
\end{equation} }
\noindent 
where $\rho$, $\mathbf{u}$, $\mathbf{B}$, $p_{\perp,\parallel}$ are the 
macroscopic variables
density, velocity, magnetic field, and thermal pressures perpendicular/parallel to the local magnetic field, respectively; 
$e = p_{\perp} + p_{\parallel}/2 + \rho u^{2}/2 + B^{2}/8\pi$ 
(for the two-temperature models CGL and BA) 
is total energy density.
For the {\it MHD} model, $e = 3p/2 + \rho u^{2}/2 + B^{2}/8\pi$.
$\Pi_P$ and $\Pi_B$ are the thermal pressure  and magnetic stress tensors, respectively,  defined by
$\Pi_{P} = p_{\perp} \mathbf{I} + (p_{\parallel} - p_{\perp}) \mathbf{bb}$ for the two-temperature models 
and 
simply $\Pi_{P} = p \mathbf{I}$ for the {\it MHD} model,  
$\Pi_{B} = (B^{2}/8 \pi) \mathbf{I} - \mathbf{BB} /4 \pi$,
where $\mathbf{I}$ is the unitary dyadic tensor and $\mathbf{b} = \mathbf{B} / B$. 
An ideal equation of state relates each temperature with its respective pressure component, and an adiabatic exponent 
$\gamma = 5/3$ is used for the MHD models.
In the source terms, $\mathbf{f}$ represents an external bulk force responsible for driving the turbulence, 
$\dot{w}$ gives the rate of change of the internal energy $w = (p_{\perp}+p_{\parallel}/2)$ of 
the gas due to heat conduction and radiative cooling, and $\dot{A}_{S}$ gives the rate of change of 
$A$ due to the microscale instabilities~\footnote{Though the physical process relaxing the macroscopic 
temperature anisotropy is 
attributed to the ions anomalous scattering in this approach, it can also 
represent (with some limitations) the situation when the relaxation is not mediated by the 
instantaneous break of magnetic momentum, as it is the case of the mirror instability development under 
continuous driving of temperature anisotropy  
(\citealt{kunz_etal_2014, riquelme_etal_2015, rincon_etal_2015, melville_etal_2016}). See discussion in 
\citet{santos-lima_etal_2016}.}

The turbulence is injected by adding a random (but solenoidal) velocity field 
(delta correlated in time)
to the gas at the end of 
each time-step. This velocity field is concentrated inside a spherical shell 
in the Fourier space of radius $k=2.5$ (i.e, with characteristic wavelength 
$L_{turb} = L/2.5$, being $L$ is the side of the cubic domain).
We employed an artificial but simple thermal relaxation prescription, 
which brings the specific internal energy $w^*$
to its initial value $w_0^*$ at a rate $\nu_{th} = 5$ 
(in code units, which gives a characteristic time approximately 20 times faster than 
the turbulence turn-over time $\tau_{turb}$)
for the models presented in Table~\ref{tab:models}:
\begin{equation}
\dot{w} = - \nu_{th} (w^* - w_0^*) \rho.
\end{equation}
The instantaneous anisotropy relaxation (represented by the source term $\dot{A}$) is implemented as follows:
after the numerical integration of the equations at each time-step,
  the anisotropy is replaced, at each grid cell,
by the marginally stable value, 
whenever this evolves to an unstable value 
beyond the threshold for the firehose
or mirror instability 
(see more details regarding the source terms and the numerical methods employed in the simulations in 
\citetalias{santos-lima_etal_2014}). As the results presented in this work 
are dimensionless and the above equations do not 
carry any physical constant, it is not necessary to attribute physical dimensions to the models.

We did not use explicit viscous or resistive terms in the numerical 
simulations (except for a small resistivity that provides a dissipation very close to the numerical one  in the {\it CGL1} model for numerical stability purposes) 
aiming at  reducing the dissipation to the minimum value provided by the numerical scheme, 
in order to maximize the inertial range of the turbulence. For the methods used in these simulations, 
the dissipation range starts at scales of approximately 16 cells (inferred from the magnetic and velocity 
power spectra of the MHD simulations).
Therefore, we cannot assess the dependence of the results with the Reynolds $R$ and/or the magnetic Prandtl
number $Pm$ ($R \equiv L_{turb} U_{turb} / \nu$ and $Pm \equiv \eta/\nu$, where $\nu$ and $\eta$ are the 
viscous and magnetic diffusivities, respectively). 
We estimate the $Pm$ number as approximately equal to unit in all our simulations.

\section{Synthetic Faraday rotation maps}

\begin{figure*}
\begin{tabular}{c c c}
	\input{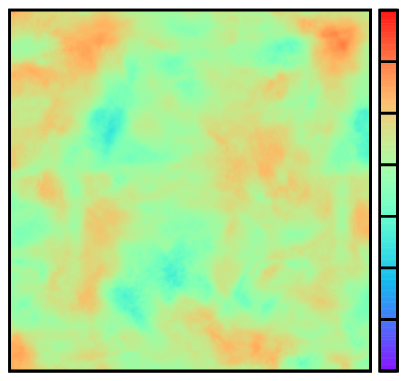} &
	\input{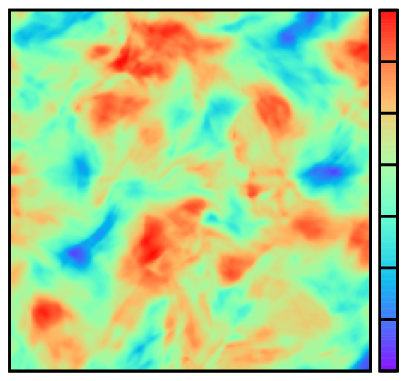} &
	\input{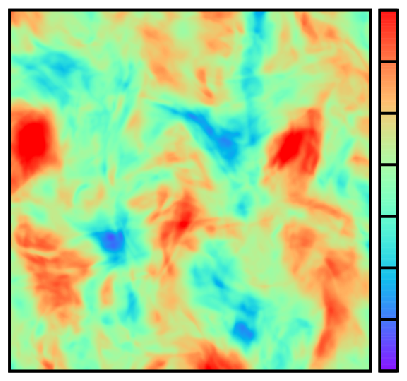}
\end{tabular}	
\caption{Normalized RM maps calculated from the simulated cubes: {\it CGL1} model with no anisotropy relaxing by the 
microscale instabilities  (left), 
{\it BA1} with fast anisotropy relaxing by the instabilities (middle), and collisional {\it MHD1} 
 model (right).
The angle $\theta$ between the line-of-sight and the direction of the uniform magnetic field is $45\degree$
for all the maps.}
\label{fig:maps}
\end{figure*}

\begin{figure}
\begin{tabular}{c}
	\input{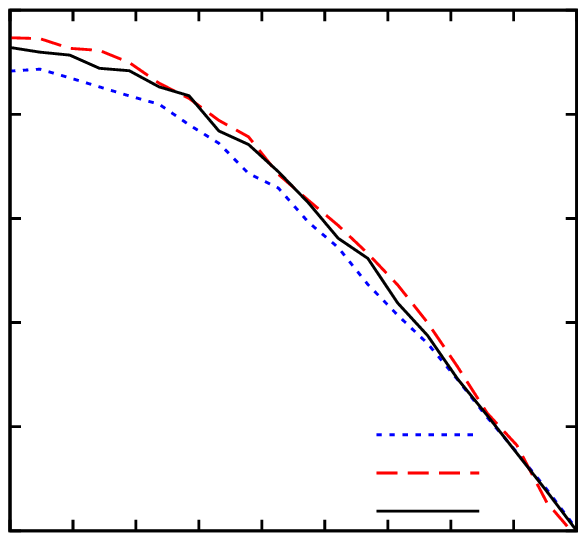} \\
	\input{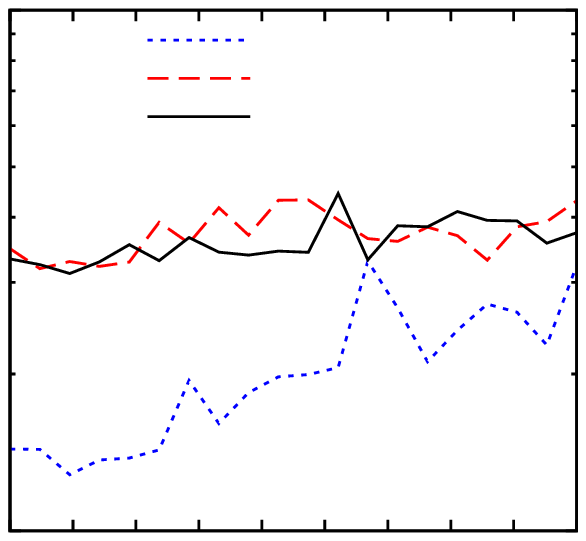}
\end{tabular}	
\caption{Normalized values of average (top) and dispersion (bottom) of the RM as a function of the angle $\theta$
between the line-of-sight and the uniform magnetic field.}
\label{fig:statistics}
\end{figure}

\begin{figure}
\begin{tabular}{c}
	\input{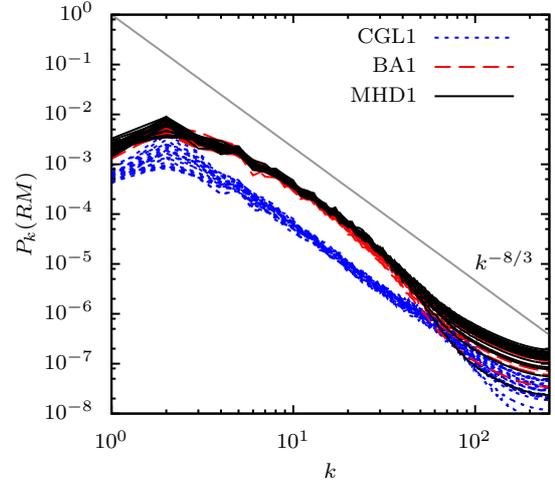}
\end{tabular}	
\caption{Power spectrum of the RM maps. The multiple lines presented for each model correspond to the power spectrum calculated 
for different angles $\theta$ between the line-of-sight and the uniform magnetic field, 
from $\theta = 0$ to $\theta = 90\degree$. A thin grey straight line with slope $-8/3$ is drawn for comparison.}
\label{fig:ps}
\end{figure}

\begin{table*}
	\centering
	\caption{Statistical moments of the synthetic RM maps built from the strongly super-Alfvenic models.}
	\label{tab:moments}
	\begin{tabular}{lccc}
		\hline
		run & 
                $\langle \delta RM^2 \rangle / \left( n_{e0} B_{rms}^2 L \right)^2$ & 
                $\langle \delta RM^3 \rangle / \langle \delta RM^2 \rangle^{3/2}$ & 
                $\langle \delta RM^4 \rangle / \langle \delta RM^2 \rangle^{2}$ \\
		\hline
		CGL2 & 
                $1.6 \times 10^{-2}$ & 
                $0.93$ & 
                $7.1$ \\
		CGL3 &
                $2.5 \times 10^{-2}$ & 
                $0.42$ & 
                $8.2$ \\
		BA2 &
                $4.0 \times 10^{-2}$ & 
                $3.6 \times 10^{-2}$ & 
                $4.0$ \\
		MHD2 &
                $3.6 \times 10^{-2}$ & 
                $-8.5 \times 10^{-3}$ & 
                $4.0$ \\
		\hline
	\end{tabular}
\end{table*}

\begin{figure}
\begin{tabular}{c}
	\input{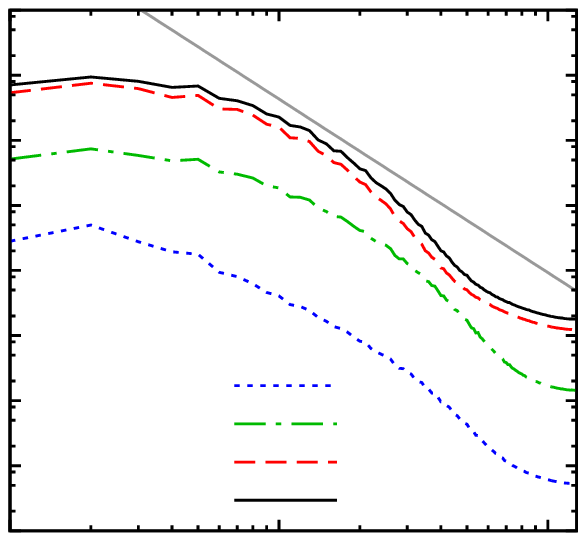}
\end{tabular}	
\caption{Same as  in Figure~\ref{fig:ps}, but for models $CGL2$, $CGL3$, $BA2$, and $MHD2$.}
\label{fig:models2}
\end{figure}

The statistics of the turbulence (that is, one and two point statistics) of 
the models described in the previous Section 
was studied in detail in \citetalias{santos-lima_etal_2014} where  models 
$Amhd$, $A2$, $A1$, $Cmhd$, $C2$, $C3$, and $C1$ 
correspond to  {\it MHD1, CGL1, BA1, MHD2, CGL2, CGL3}, and  $BA2$,
respectively. 

Figure~\ref{fig:maps} presents the maps of the dimensionless Faraday rotation measurement ($RM$):
\begin{equation}
RM = \int_0^L n_e B_{LOS} dl,
\end{equation}
normalized by $n_{e0} B_0 L$ (where $B_0$ is the intensity of the mean magnetic field, $n_{e0}$ is the 
average density of electrons and $L$ is the length of the Faraday screen) for 
the models {\it CGL1, BA1}, and {\it MHD1}, 
calculated for an arbitrary line-of-sight (LOS) 
whose direction has angle $\theta = 45 \degree$ with the mean magnetic field. The last snapshot 
of the simulations (at $\approx 10 \tau_{turb}$)
were used for the calculations.

A visual inspection shows that the RM map of model {\it CGL1} presents fluctuations of smaller amplitude
compared to the {\it MHD1} model. Model {\it BA1} on the other hand, has the RM map appearance similar 
to model {\it MHD1}.

Figure~\ref{fig:statistics} shows the normalized values of the average  (top) 
and dispersion (bottom) of RM as a function of the angle $\theta$, for the mildly super-Alfvenic models. The maps were built using 20 values of $\theta$ equally 
spaced between $\theta = 0$ and $90\degree$, and the statistical moments were averaged over maps built 
from the different 
snapshots available for each model. 
The two-temperature 
models develop excess of perpendicular pressure
in most of the domain ($A>1$), and larger anti-correlation between 
the magnetic and density fluctuations (when compared to the one-temperature collisional MHD model; 
see Figure 10 in \citetalias{santos-lima_etal_2014}). This enhanced anti-correlation is expected to lead to a net 
reduction of the rotation  measure in the case of the {\it CGL1} model when compared to the 
{\it MHD1} model. However, 
this reduction is found  to be small (only a few percent) for small angles $\theta$ in the top plot 
of Figure~\ref{fig:statistics}. For increasing angles $\theta$, the mean RM for the {\it CGL1} 
model 
converges to values close to  the MHD model. The model {\it BA1} has the RM mean value
 similar to the {\it MHD1} model for all angles.

Due to the dominance of the perpendicular temperature component in the {\it CGL1} model, 
the thermal stresses offer resistance to motions perpendicular to the local field lines then reducing 
the fluctuations of the magnetic fields.
In consequence, the fluctuations 
of the RM produced by the magnetic field turbulence are also affected. The bottom panel 
of Figure~\ref{fig:statistics} compares the normalized dispersion of  RM  for the three models. This relative dispersion 
of the {\it CGL1} is about 2 times smaller (for small $\theta$) compared to the {\it MHD1} model, 
and this difference is smaller for larger values of $\theta$. The inclusion of the fast anisotropy relaxation by the 
microscale instabilities
(model {\it BA1}) makes this relative dispersion in  RM  very similar to the MHD
model.

Figure~\ref{fig:ps}  compares the power spectrum of the RM maps for the 
mildly super-Alfvenic models. For each model,
 the power spectrum is shown for different values of $\theta$ (from $\theta = 0$ to $90\degree$).
For wavenumbers approximately in the estimated inertial range ($5<k<30$) the slopes of the power 
spectrum for the different lines of sight  are nearly the same.
While the {\it BA1} model has almost  indistinguishable power spectrum from the {\it MHD1} model, the 
model {\it CGL1} has  less power in all  scales and is slightly  flatter. 
We also note that the RM spectrum of the {\it CGL1} model  has a power law close to $k^{-8/3}$ 
(which is expected 
when only the magnetic field fluctuates, that is, the density fluctuations are negligible) and  an unidimensional 
power spectrum $|B_k|^2 \propto k^{-5/3}$, 
while model {\it MHD1}  has a slightly flatter slope at small $k$ values and then becomes slightly steeper at larger $k$. 
The corresponding  magnetic power spectrum 
is slightly flatter than the Kolmogorov power law 
$k^{-5/3}$ in our simulation as shown in Fig. 6 of \citetalias{santos-lima_etal_2014}; 
the same can be observed in a similar simulation presented in \citetalias{nakwacki_etal_2016} 
(Fig. 6 left panel, model Bext=1, cs=1).
Compared to model {\it MHD1}, the slow decay of the dissipation range of model {\it CGL1}  
points to an accumulation of power at the small scales
(which are not properly solved by our grid resolution) caused by the kinetic instabilities (\citetalias{nakwacki_etal_2016}).

We repeated the  analysis above for the strongly super-Alfvenic models 
{\it CGL2, CGL3, BA2}, and {\it MHD2}. 
Naturally, for these models the turbulent component of the magnetic field dominates the uniform one. This implies that  
 the statistics of the RM maps built from these models 
is generally independent of the adopted LOS (except for the value corresponding to the mean field).  In fact the statistics 
of the CGL  models  RM maps keeps a marginal dependence on the LOS, as the turbulent magnetic field is not as amplified here as  in the MHD case.
Table~\ref{tab:moments} shows the statistical moments for  RM  averaged over maps with different LOS 
(using 20 values of $\theta$ uniformly spaced in the interval between $0$ and $90 \degree$).
The dispersion values $\langle \delta RM^2\rangle$ are compared to $(n_{e0} B_{rms} L)^2$, in order 
to check how precisely we can track the intensity of the turbulent component of the magnetic field.
Compared to the MHD model, the CGL models give a smaller value (by a factor of two), but this  also depends 
on the compressibility of the turbulence, being slightly higher for the more compressible model $CGL3$.
The model with bounds on the anisotropy $BA2$ shows a dispersion similar to the MHD case.
The skewness and kurtosis of the distribution of the RM  are also shown in Table~\ref{tab:moments}
(in columns $\langle \delta RM^3 \rangle / \langle \delta RM^2 \rangle^{3/2}$ and 
$\langle \delta RM^4 \rangle / \langle \delta RM^2 \rangle^2$, respectively). While model {\it BA2} presents results very 
similar to  {\it MHD2}, with nearly zero skewness and the same values for the kurtosis, the 
CGL models show a positive skewness (which means a longer tail of large values) and a kurtosis 
approximately twice that of the MHD model, so that the distribution being is more peaked. 

Figure~\ref{fig:models2} shows the power spectrum of RM for the highly super-Alfvenic models (averaged over the 
different LOS).
The curves for the CGL models are displaced in the vertical axis and the values are multiplied by 
a factor of $100$ in order to make the difference of the slopes between the models better observed.
Similar to the mildly super-Alfvenic case, the CGL models present 
a flatter spectrum  at large $k$ values of the inertial range compared to the standard MHD model. The slope for model $CGL2$ is even flatter than $k^{-8/3}$
(due to the increase of the small scale magnetic fluctuations caused by the instabilities which are stronger 
in this high beta plasma regime compared to the previous mildly super-Alfvenic case), while the {\it MHD2} 
model exhibits a power similar to the mildly super-Alfvenic case.

\section{Summary and conclusions}

In this work we explored the role of plasma collisionless effects on simulated Faraday rotation maps resembling the conditions of 
the intracluster medium of galaxies (ICM). We presented a
statistical analysis of the Faraday rotation maps obtained from  simulations of forced turbulence  
in a three-dimensional domain with periodic boundaries considering  three different models of the ICM plasma. The first 
one-temperature collisional MHD 
model  considers isotropy in the velocity thermal distribution of the particles, 
an assumption  that is not suitable  {\it a priori} for the weakly collisional ICM, 
where the mean free path for  ion-ion Coulomb collisions is distributed typically in the range $2-100$~kpc
(see \citealt{egan_etal_2016}). The second model (CGL) allows for
the development of anisotropy in the velocity thermal distribution (two-temperature approach), 
according to the conservation of the first adiabatic invariant (the magnetic momentum) of charged 
particles and the absence of heat conduction. The third model 
(BA) differs from the second by the inclusion of  
 a phenomenological constraint on the temperature anisotropy due to the fast development
 of the firehose and mirror instabilities at the microscales (much smaller than the typical turbulence scales, 
 reaching the ions kinetic scales).
 These instabilities are triggered by the 
temperature anisotropy itself. 

Compared to the Faraday rotation maps resulting from the one-temperature collisional model (MHD), 
those from the collisionless CGL model
present a relative dispersion smaller, with a steeper and less intense power 
spectrum  in all  scales. 
On the other hand, the statistical properties  of the RM  maps resulting from the collisionless BA model, 
which bounds the anisotropy to the firehose and mirror stable thresholds,
are very similar to those of the MHD  model. 
As stressed in Section 1, in \citetalias{nakwacki_etal_2016} we performed a similar RM analysis of collisionless 
two-temperature (with fixed values) models for the intracluster medium, but without considering the effects of the 
thermal relaxation by the kinetic instabilities. In this case the results were 
similar to those of the CGL model  above, i.e., with significant differences in the RM maps and 
their statistical properties with regard to the collisional MHD model. Specifically,   important  
imprints of the pressure anisotropy were found to prevail in the magnetic field structure  resulting  
in Faraday rotation maps with smaller correlation lengths.

It has been demonstrated in \citetalias{santos-lima_etal_2014} that the inclusion of the 
anisotropy relaxation by the kinetic mirror and firehose instabilities in collisionless 
two-temperature systems makes the statistical properties  of the turbulence (in high $\beta$ plasmas) 
as well as the amplification of the  magnetic fields  via the small-scale turbulent dynamo very similar 
to those of  collisional MHD systems. The later approach is in fact  used  in most numerical 
simulations of the intracluster medium. 

Therefore, the present result, {\it in principle}, reinforces the justification for the use of the collisional MHD approximation at least in studies of the large scale properties of the ICM. 
Nevertheless, this study has limitations and several questions still remain opened,  as we briefly address below.

Recently \cite{santos-lima_etal_2016} have reviewed the limitations of the 
 anisotropy relaxation approach employed, e.g., in \citetalias{santos-lima_etal_2014}.  
For instance, this neglects the effects of the
microscale magnetic fields generated by microinstabilities 
on the stretching rate of the large scale component (near the injection scale of the turbulence) (see also \citealt{schekochihin_cowley_2006, mogavero_schekochihin_2014, melville_etal_2016}).

Furthermore, the present study has focussed 
only
on the subsonic regime of the turbulence 
driven by purely solenoidal forcing, 
which explains the dominance of incompressible motions. 
On the other hand, the turbulence generated by the merging processes in the ICM 
is expected to be 
partially compressional (at the injection scales)
and at least mildly supersonic 
(\citealt{brunetti_jones_2014, bruggen_vazza_2015, bykov_etal_2015}) 
and, in fact, a compressible cascade  in the ICM can reach small scales ($0.1-1$ kpc) 
before being dissipated. 
This  implies that the magnetic fields can be entangled and/or advected also by compressive motions. In addition, weak shocks and collisionless effects will also affect the microphysics of processes like  heating transport and thermal conduction (e.g., \citealt{santos-lima_etal_2016}), and
may  be important to the re-acceleration of particles in the ICM (see for example \citealt{brunetti_lazarian_2007, brunetti_lazarian_2011}).
The complex interplay between  compressible modes (and shocks) and collisionless effects 
(as the collisionless damping) which have been neglected in the present  collisionless MHD approach turn it inadequate to treat the compressible turbulent regime of the ICM
(see further discussion on this subject in \citealt{santos-lima_etal_2016}).

\section*{Acknowledgements}

RSL acknowledges support from a grant of the Brazilian Agency FAPESP (2013/15115-8), 
EMGDP partial support from  FAPESP (2013/10559-5) and CNPq (306598/2009-4) grants.
G.K. acknowledges support from FAPESP (grants no. 2013/04073-2 and 2013/18815-0) and PNPD/CAPES (grant no. 1475088) through a Postdoctoral Fellowship at University Cruzeiro do Sul.
The numerical simulations in this work were carried out in the supercluster of the Astrophysical Informatics Laboratory (LAi) of IAG-USP and UnicSul whose purchase was made possible by FAPESP.
The authors would also like to acknowledge the anonymous referee 
for the critics and suggestions which helped to improve this work.





\begin{thebibliography}{99}



\bibitem[\protect\citeauthoryear{Beresnyak 
\& Miniati}{2016}]{beresnyak_miniati_2016} Beresnyak A., Miniati F., 2016, ApJ, 817, 127 


\bibitem[\protect\citeauthoryear{Br{\"u}ggen \& Vazza}{2015}]{bruggen_vazza_2015} Br{\"u}ggen M., Vazza F., 2015, ASSL, 407, 599

 
\bibitem[\protect\citeauthoryear{Brunetti \& Jones}{2014}]{brunetti_jones_2014} Brunetti G., Jones T.~W., 2014, IJMPD, 23, 1430007-98 


\bibitem[\protect\citeauthoryear{Brunetti \& Lazarian}{2011}]{brunetti_lazarian_2011} Brunetti G., Lazarian A., 2011, MNRAS, 412, 817 


\bibitem[\protect\citeauthoryear{Brunetti \& Lazarian}{2007}]{brunetti_lazarian_2007} Brunetti G., Lazarian A., 2007, MNRAS, 378, 245 


\bibitem[\protect\citeauthoryear{Bykov et al.}{2015}]{bykov_etal_2015} Bykov A.~M., Churazov E.~M., Ferrari C., Forman W.~R., Kaastra J.~S., Klein U., Markevitch M., de Plaa J., 2015, SSRv, 188, 141 


\bibitem[\protect\citeauthoryear{Chew, Goldberger, 
\& Low}{1956}]{chew_etal_1956} Chew G.~F., Goldberger M.~L., Low F.~E., 1956, RSPSA, 236, 112 


\bibitem[\protect\citeauthoryear{Egan et al.}{2016}]{egan_etal_2016} 
Egan H., O'Shea B.~W., Hallman E., Burns J., Xu H., Collins D., Li H., 
Norman M.~L., 2016, arXiv, arXiv:1601.05083 


\bibitem[\protect\citeauthoryear{En{\ss}lin, Vogt, 
\& Pfrommer}{2005}]{ensslin_etal_2005} En{\ss}lin T., Vogt C., Pfrommer C., 2005, mpge.conf, 231 


\bibitem[Falceta-Gon{\c c}alves \& Kowal(2015)]{falceta15} Falceta-Gon{\c c}alves, D., \& Kowal, G.\ 2015, \apj, 808, 65 


\bibitem[\protect\citeauthoryear{Feretti et al.}{1995}]{ferreti_etal_1995} Feretti L., Dallacasa D., Giovannini G., Tagliani A., 1995, A\&A, 302, 680 


\bibitem[\protect\citeauthoryear{Gary}{1993}]{gary_1993} Gary 
S.~P., 1993, tspm.book, 193 


\bibitem[\protect\citeauthoryear{Kotarba et 
al.}{2011}]{kotarba_etal_2011} Kotarba H., Lesch H., Dolag K., Naab T., 
Johansson P.~H., Donnert J., Stasyszyn F.~A., 2011, MNRAS, 415, 3189 


\bibitem[\protect\citeauthoryear{Kowal, Falceta-Gon{\c c}alves, 
\& Lazarian}{2011}]{kowal_etal_2011} Kowal G., Falceta-Gon{\c c}alves D.~A., Lazarian A., 2011, NJPh, 13, 053001 


\bibitem[\protect\citeauthoryear{Kunz, Schekochihin, 
\& Stone}{2014}]{kunz_etal_2014} Kunz M.~W., Schekochihin A.~A., Stone J.~M., 2014, PhRvL, 112, 205003 


\bibitem[\protect\citeauthoryear{Melville, Schekochihin, \& Kunz}{2016}]{melville_etal_2016} Melville S., Schekochihin A.~A., Kunz M.~W., 2016, MNRAS, 459, 2701 


\bibitem[\protect\citeauthoryear{Mogavero \& Schekochihin}{2014}]{mogavero_schekochihin_2014} Mogavero F., Schekochihin A.~A., 2014, MNRAS, 440, 3226 


\bibitem[\protect\citeauthoryear{Nakwacki et 
al.}{2016}]{nakwacki_etal_2016} Nakwacki M.~S., Kowal G., Santos-Lima R., 
de Gouveia Dal Pino E.~M., Falceta-Gon{\c c}alves D.~A., 2016, MNRAS, 455, 
3702 


\bibitem[\protect\citeauthoryear{Rincon, Schekochihin, 
\& Cowley}{2015}]{rincon_etal_2015} Rincon F., Schekochihin A.~A., Cowley S.~C., 2015, MNRAS, 447, L45 


\bibitem[\protect\citeauthoryear{Riquelme, Quataert, 
\& Verscharen}{2015}]{riquelme_etal_2015} Riquelme M.~A., Quataert E., Verscharen D., 2015, ApJ, 800, 27 


\bibitem[\protect\citeauthoryear{Santos-Lima et 
al.}{2014}]{santos-lima_etal_2014} Santos-Lima R., de Gouveia Dal Pino E.~M., 
Kowal G., Falceta-Gon{\c c}alves D., Lazarian A., Nakwacki M.~S., 2014, 
ApJ, 781, 84 


\bibitem[\protect\citeauthoryear{Santos-Lima et al.}{2016}]{santos-lima_etal_2016} Santos-Lima R., Yan H., de Gouveia Dal Pino E.~M., Lazarian A., 2016, MNRAS, 460, 2492 


\bibitem[\protect\citeauthoryear{Schekochihin 
\& Cowley}{2006}]{schekochihin_cowley_2006} Schekochihin A.~A., Cowley S.~C., 2006, PhPl, 13, 056501 


\bibitem[\protect\citeauthoryear{Sharma et al.}{2006}]{sharma_etal_2006} Sharma P., Hammett G.~W., Quataert E., Stone J.~M., 2006, ApJ, 637, 952 


\bibitem[\protect\citeauthoryear{Sironi}{2015}]{sironi_2015} Sironi 
L., 2015, ApJ, 800, 89 


\bibitem[\protect\citeauthoryear{Sironi 
\& Narayan}{2015}]{sironi_narayan_2015} Sironi L., Narayan R., 2015, ApJ, 800, 88 



\end{thebibliography}




%
%



\bsp	
\label{lastpage}
\end{document}